\documentclass[twocolumn,pra,amsmath]{revtex4-1}
\usepackage{epsfig,bm}
\begin{document}

\author{Kui-Tian Xi$^1$, Jinbin Li$^{1,2,}$}\email{jinbin@nuaa.edu.cn}
\author{Da-Ning Shi$^{1,}$}
\affiliation
{$^1$College of Science, Nanjing University of Aeronautics
and Astronautics, Nanjing, 211106, People's Republic of China \\
$^2$Kavli Institute for Theoretical Physics China, CAS, Beijing, 100190, People's Republic of China}

\title{Phase separation of a two-component dipolar Bose-Einstein condensate in the quasi-one-dimensional and quasi-two-dimensional regime}

\begin{abstract}
We consider a two-component Bose-Einstein condensate, which contains atoms with magnetic dipole moments aligned along the $z$ direction (labeled as component 1) and nonmagnetic atoms (labeled as component 2). The
problem is studied by means of exact numerical simulations. The
effects of dipole-dipole interaction on phase separations are
investigated. It is shown that, in the quasi-one-dimensional regime,
the atoms in component 2 are squeezed out when the dimensionless
dipolar strength parameter is small, whereas the atoms in component
1 are pushed out instead when the parameter is large. This is
in contrast to the phenomena in the quasi-two-dimensional regime.
These two components are each kicked out by
the other in the quasi-one-dimensional regime and this phenomenon is discussed as well.
\end{abstract}

\maketitle

\section{Introduction}\label{sec:introduction}

Over the past decade, the experimental realization of a Bose-Einstein
condensate (BEC) of $^{52}$Cr atoms
\cite{Pfau2002,Pfau2005,Pfau2006,Pfau2007} with a strong dipole-dipole
interaction has given new impetus to theoretical investigations of
BECs with long-range interaction at low temperatures. Due to the large
magnetic dipole moment of Cr atoms \cite{Pfau2005,Pfau2006},
$\mu_{Cr}=6\mu_{B}$ (where $\mu_{B}$ is the Bohr magneton),
the anisotropic interaction between polarized magnetic dipoles results
in anisotropic deformations during expansion of the condensate.
Dipolar quantum gases are governed by the $d$-wave symmetry of the
long-range dipole-dipole interaction, which causes novel properties
such as unusual stability properties \cite{Pfau2000}, exotic ground
states \cite{Dell2002,Yi2000}, and modified excitation spectra
\cite{Yi200102,Santos2002_2}. For a review of dipolar condensates, see \cite{Lahaye2009}.

The miscibility and separation of a two-component BEC have been studied
theoretically \cite{Ho,Pu} and experimentally \cite{Cornell1997} previously. Recently, more detailed and controlled experimental results have been obtained, illustrating the effects of phase
separation in a multi-component BEC \cite{Cornell2004}. In all these
papers, the studies of the binary condensates were limited to the
case of $s$-wave interactions; while a great deal of attention has
been drawn recently to dipolar BECs. Hexagonal, labyrinthine,
solitonlike structures, and hysteretic behavior have been studied in
a two-component dipolar BEC \cite{Ueda2009}, as well as
immiscibility-miscibility transitions \cite{Malomed2010}.

The theoretical description of a BEC in the dilute limit in the
framework of the extended Gross-Pitaevskii equation (GPE), which has
been part of many text books on quantum mechanics or Bose-Einstein
condensates \cite{Pethick}, is well known. The time-independent
GPE reads
\begin{equation}\label{eq:dipolarGPE}
\left[-\frac{\hbar^2}{2m}\nabla^2+V(\bm r)+UN|\psi(\bm
r)|^2+\Phi_{dd}(\bm r)\right]\psi(\bm r)=\mu\psi(\bm r),
\end{equation}
where $\mu$ is the chemical potential, $m$ is the atomic mass, $n(\bm r)=N|\psi(\bm r)|^2$ is the density
of atoms, which is normalized via $\int n(\bm
r)d^{3}{\bm r}=N$, and $V(\bm r)=m\left(\omega_x^2x^2+\omega_
y^2y^2+\omega_z^2z^2\right)/2$ is the external harmonic trap confining the gas.
The local mean-field potential $U|\psi(\bm r)|^2$ represents the
$s$-wave interaction, with $U=4\pi\hbar^2a_s/m$ and
scattering length $a_s$. The long-range interaction between two
particles, $\Phi_{dd}(\bm r)$, represents the dipolar interaction,
which is given by \cite{Yi2000,Yi200102}
\begin{equation}\label{eq:Phidd}
\Phi_{dd}(\bm r)=\int d^{3}{\bm r}' U_{dd}\left({\bm r}-{\bm
r}'\right)|\psi\left({\bm r}'\right)|^2.
\end{equation}
Here $U_{dd}({\bm r})=\left[C_{dd}/\left(4\pi\right)\right]\hat{e}_i\hat{e}_{j}
\left(\delta_{ij}-3\hat{r}_i \hat{r}_{j}\right)/r^{3}$ describes
the interaction between two dipoles which are aligned by an external
field along a unit vector $\hat{{\bm e}}$ through a distance ${\bm r}$. The coupling $C_{dd}=E^{2}\alpha^{2}/\epsilon_{0}$ where the dipoles are induced by an electronic field $\bm E =E\hat{\bm e}$, with the static polarizability $\alpha$ and the permittivity of free space $\epsilon_{0}$. If the atoms have a magnetic dipole moment $d_{m}$ aligned by a magnetic field $\bm B=B\hat{\bm e}$, the coupling is $C_{dd}=\mu_{0}d_{m}^{2}$ with the permeability of free space $\mu_{0}$ \cite{Dell20040507}.  A measurement of the strength of the dipole-dipole interaction
relative to the $s$-wave scattering energy is provided by a
dimensionless quantity (the so-called dimensionless
dipole-dipole strength parameter) $\varepsilon_{dd}\equiv
C_{dd}/\left(3U\right)$. For dipoles aligned along $\bm z$, this dipolar
mean-field potential (\ref{eq:Phidd}) can be expressed in
terms of a fictitious electrostatic potential $\phi(\bm
r)$ \cite{Dell20040507}
\begin{equation}\label{eq:rwPhidd}
\Phi_{dd}(\bm r)=-U\varepsilon_{dd}\left[3
\partial_{z}^2\phi(\bm r)+ n(\bm r)\right],
\end{equation}
where $n(\bm r)$ and $\phi(\bm r)$ satisfy Poisson's equation
$\nabla^2\phi=-n(\bm r)$. This formulation of the problem allows us
to immediately identify some generic features of dipolar gases. For
example, if $n(\bm r)$ and hence $\phi(\bm r)$ is uniform along the
polarization direction $\bm z$, then the nonlocal part of the dipolar
interaction vanishes because of the operator $\partial^2_z$. The details are
in Sec.\ref{sec:twodregime}.

The two-component dipolar BEC, confined in a cylindrical trap, is
described by two coupled Gross-Pitaevskii equations. We take Cr as
component 1 and Rb as component 2, then the GP equations can be written as
\begin{eqnarray}
&&H_i\psi_i(\bm r)=\mu_i\psi_i(\bm r)\\
H_1&=&-\frac{\hbar^2}{2m_1}\nabla^2+V_1(\bm r)+U_1N_1|\psi_1
(\bm r)|^2+\nonumber\\&&U_{12}N_2|\psi_2(\bm r)|^2+\Phi_{dd}(\bm r) \label{eq:h1}\\
H_2 &=& -\frac{\hbar^2}{2m_2}\nabla^2+V_2(\bm r)+U_2N_2|\psi_2
(\bm r)|^2 \nonumber\\
& & +U_{12}N_1|\psi_1(\bm r)|^2, \label{eq:h2}
\end{eqnarray}
where $V_i(\bm r)=m_i\left(\omega_{i\rho}^2\rho^2+
\omega_{iz}^2z^2\right)/2$ $(i=1,2)$ represents a cylindrical harmonic trap
with radial trap frequency $\omega_{i\rho}$ and axial trap frequency
$\omega_{iz}$ for the $i$th component, and $\rho=\sqrt{x^2+y^2}$.
$\psi_i(\bm r)=\psi_i(\rho)\psi_i(z)$ is the wave function of
component $i$ with particle number $N_i$. $m_i$ and $\mu_i$ are the mass
and the chemical potential for the $i$th component, respectively.

The interatomic and the intercomponent $s$-wave scattering interactions are described by $U_i$ and $U_{12}$, respectively, with the following expressions \cite{Pethick}:
\[U_i=\frac{4\pi\hbar^2a_i}{m_i},~U_{12}=\frac{2\pi\hbar^2a_{12}}{m_1m_2/\left(m_1+m_2\right)},\]
where $a_i$ is the scattering length of component $i$ and $a_{12}$ is
that between component 1 and 2.

The goal of this paper is to analyze the properties of phase separation
in a two-component dipolar BEC in quasi-one and quasi-two dimensions,
and it is organized as follows. In Sec. \ref{sec:onedregime},
the Gross-Pitaevskii equations of the two-component dipolar BEC are
reduced to quasi-one-dimensional form. Analytical insights into the
phenomenology of phase separation in the axial direction are shown. In
Sec. \ref{sec:twodregime}, the quasi-two-dimensional Gross-Pitaevskii
equations representing the two-component dipolar BEC are derived,
and analytical results for the phase separation in the radial direction are
given. Finally, the paper is concluded in Sec.\ref{sec:conclusion}.

\section{Quasi-one-dimensional Regime (infinite pancake)}\label{sec:onedregime}
We begin by neglecting the radial trapping frequency ($\omega_{i\rho}=
0$) and increase the axial trapping frequency ($\omega_{iz}$) to the extent that a highly flattened pancake-shaped BEC is produced with infinite radial extent. In this case, the trap has infinite radial extent and uniform two-dimensional (2D) density in the $\rho$ direction.
The Poisson equation $\nabla^2\phi=-n(\bm r)$ reduces to
$\partial_{z}^2\phi=-n\left(z\right)$, and Eq. \eqref{eq:rwPhidd}
reduces to the contactlike form \cite{Dell2008}
\begin{equation}
\Phi_{dd}(\bm r)=2U_1\varepsilon_{dd}n\left(z\right)=
2U_1N_1\varepsilon_{dd} |\psi_1\left(z\right)|^2.
\end{equation}
If we let $\sqrt{\hbar/\left(m_1\omega_{1z}\right)}$ and
$\hbar\omega_{1z}$ be the units for length and energy, respectively,
then we can rewrite Eqs. \eqref{eq:h1} and \eqref{eq:h2} as two
dimensionless coupled differential equations:
\begin{eqnarray}
H_1&=&-\frac{1}{2}\frac{\partial^2}{\partial z^2}+\frac{1}{2}z^2+4\pi N_1a_1
(1+2\varepsilon_{dd})|\psi_1(z)|^2+\nonumber\\&&\frac{1+a_{m}}{a_{m}}2\pi
N_2a_{12}|\psi_2(z)|^2,\label{eq:dlessh1}\\H_2&=&-\frac{1}{2a_{m}}\frac{\partial^2}{\partial z^2}+\frac{1}{2}a_{m}
a_{\omega}^2z^2+\frac{4\pi N_2a_2}{a_{m}}|\psi_2\left(z\right)|^2+\nonumber\\
&&\frac{1+a_{m}}{a_{m}}2\pi N_1a_{12}|\psi_1(z)|^2,
\label{eq:dlessh2}
\end{eqnarray}
where $a_{m}=m_2/m_1$ and $a_{\omega}=\omega_{2z}/\omega_{1z}$.

Using the finite-difference approximation \cite{finitediff}
\begin{equation}
\frac{\partial^2 \psi_{i}}{\partial z^2}=\frac{\psi_i^{l+1}-2\psi_i^{l}+\psi_i^{l-1}}{h^2}, \nonumber
\end{equation}
we can write $H_i$ as a symmetric tridiagonal matrix with diagonal elements $\upsilon_{i}^{l}$ and subdiagonal or superdiagonal elements $w_{i}^{l}$ in $H_{i}$, where
\begin{eqnarray}
\upsilon_{1}^{l} &=& \frac{1}{2}lh^{2}+4\pi N_{1}a_{1}\left(1+2\varepsilon\right)|\psi_{1}^{l}|^{2}+ \nonumber\\
& & \frac{1+a_{m}}{a_{m}}2\pi N_{2}a_{12}|\psi_{2}^{l}|^{2}+\frac{1}{h^{2}}, \nonumber\\
\upsilon_{2}^{l} &=& \frac{1}{2}a_{m}a_{\omega}^{2}lh^{2}+\frac{4\pi N_{2}a_{2}}{a_{m}}|\psi_{2}^{l}|^{2}+ \nonumber\\
& & \frac{1+a_{m}}{a_{m}}2\pi N_{1}a_{12}|\psi_{1}^{l}|^{2}+\frac{1}{a_{m}h^{2}}, \nonumber\\&&
w_{1}^{l}=-\frac{1}{2h^{2}},~
w_{2}^{l}=-\frac{1}{2a_{m}h^{2}}, \nonumber
\end{eqnarray}
with mesh length $h$ in the $z$ direction.
Here, free boundary conditions should be applied as $\lim\limits_{z\rightarrow\pm\infty}\psi_i(z)=0$. Diagonalizing these two symmetric tridiagonal matrices, we can obtain the ground state wave functions $\psi_{i}\left(z\right)$ and the density profiles of the dipolar BEC \cite{Pu}.

In this case, the critical value $a_{12}^{c}$ which causes phase separation is
\begin{equation}\label{eq:criterion}
a_{12}^{c}=\sqrt{\frac{4a_1\left(1+2\varepsilon_{dd}\right)a_2a_{m}}{\left(1+a_{m}
\right)^2}},
\end{equation}
arising from $U_1^{eff1}U_2-U_{12}^2=0$ with $U_1^{eff1}=U_1\left(1+2\varepsilon_{dd}\right)$. This criterion for the strength of the repulsive intercomponent interaction is independent of the atom numbers as well as of the trap strength \cite{Chui1998}.

In our calculations, we take the scattering lengths of $^{52}$Cr and
$^{87}$Rb as 5 nm \cite{Pfau2006} and 10 nm \cite{Feshbach}, respectively. For the trap, we assume
$\omega_{1z}=2\pi\times 160$ Hz and $a_{m}a_{\omega}^2=1$. In this
case, we take the particle number as $N_1=N_2=1000$.

\begin{table}
\caption{\label{table:table1} Dimensionless values of interactions in quasi-one dimension}
\begin{tabular}{cccc}
  \hline\hline
  $\varepsilon_{dd}$ & $U_{1}^{eff1}$ ($10^{-3}$) & $U_{12}$
  ($10^{-3}$) & $U_{2}$ ($10^{-3}$) \\ \hline
  -0.3 & 25 & \begin{tabular}{c}
                40 ($a_{12}=4$ nm) \\
                43 ($a_{12}^{c}=4.33$ nm) \\
              \end{tabular}
   & 75 \\
  0.1 & 75 & \begin{tabular}{c}
                75 ($a_{12}^{c}=7.5$ nm) \\
                76 ($a_{12}=7.6$ nm) \\
              \end{tabular}
   & 75 \\
  0.5 & 125 & \begin{tabular}{c}
                80 ($a_{12}=8$ nm) \\
                97 ($a_{12}^{c}=9.68$ nm) \\
              \end{tabular}
   & 75 \\
  \hline\hline
\end{tabular}
\end{table}

\begin{figure}[b]\begin{center}
\includegraphics[width=1.0\linewidth]{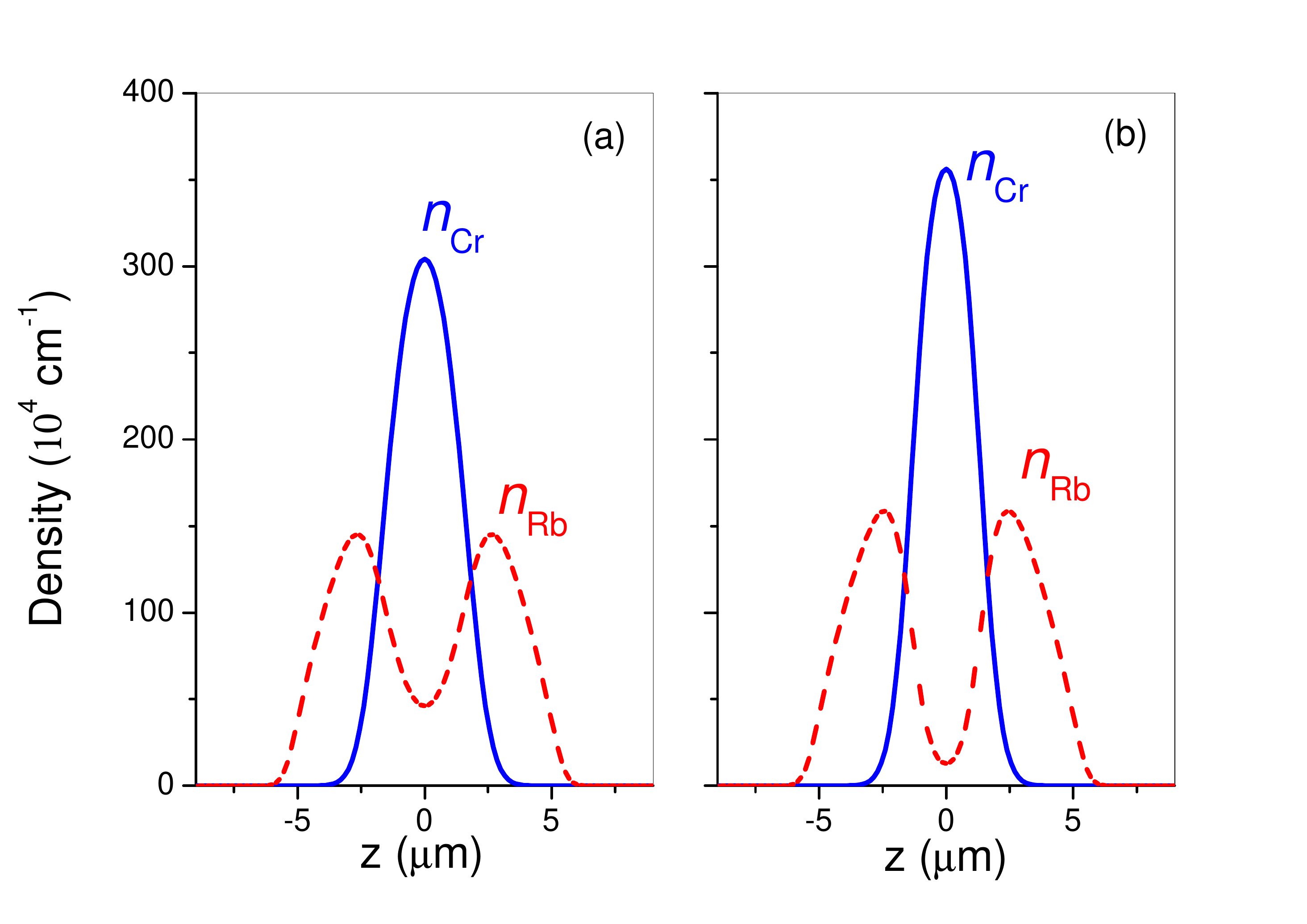}
\end{center}\vspace{-1cm}
\caption{\label{fig:Fig1} (Color online) Ground-state density profiles
in the two-component dipolar BEC for $\varepsilon_{dd}=-0.3$ and
different values of intercomponent scattering length: (A) $a_{12}=4$
nm; (B) $a_{12}=4.33$ nm (the critical
point). In this figure and Figs. \ref{fig:Fig2} and \ref{fig:Fig3}, the blue solid line represents the
density profile of $^{52}$Cr and the red dashed line is that of
$^{87}$Rb.}
\end{figure}

\begin{figure}[t]\begin{center}
\includegraphics[width=1.0\linewidth]{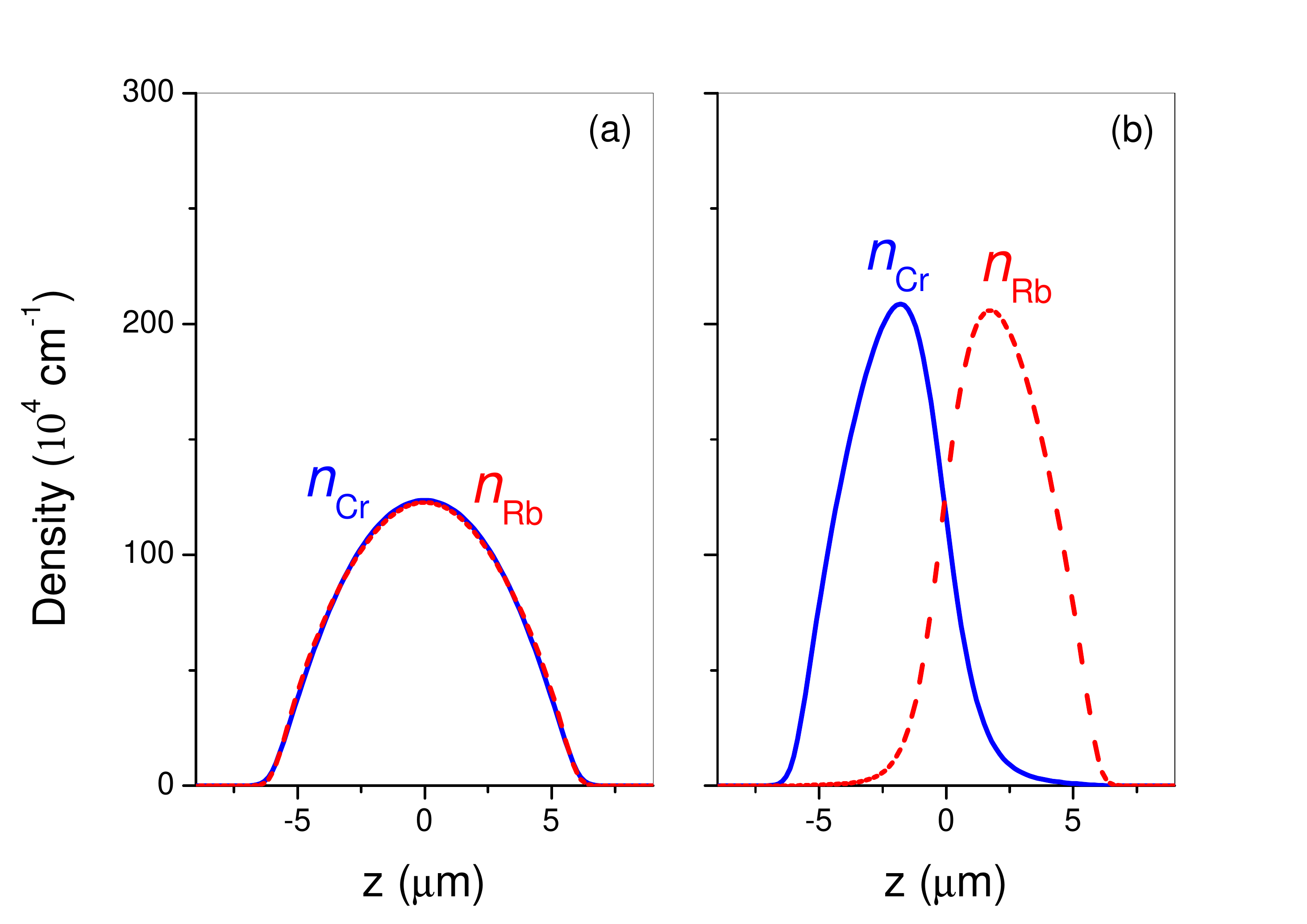}
\end{center}\vspace{-0.8cm}
\caption{\label{fig:Fig2} (Color online) Ground-state density profiles in the
two-component dipolar BEC for $\varepsilon_{dd}=0.1$ and different
values of the intercomponent scattering length: (A) $a_{12}=7.5$ nm (the
critical point); (B) $a_{12}=7.6$ nm.}
\end{figure}

First, we assume $\varepsilon_{dd}=-0.3$ and calculate the
density profiles of those two components for various $a_{12}$ using
the coupled GP equations. For a relatively small value of the
intercomponent scattering length $a_{12}$ (4 nm in Table \ref{table:table1}), the two-component BEC does
not separate [see Fig. \ref{fig:Fig1} (A)]. Then we increase
$a_{12}$ to the critical point (4.33 nm in Table \ref{table:table1}), the phase separation of Cr and Rb atoms occurs and the density of Cr increases [see Fig. \ref{fig:Fig1} (B)]. The dipoles in this case lie predominantly side by side, and the net dipolar interaction is
attractive (repulsive) when $\varepsilon_{dd}<0$ $\left(\varepsilon_{dd}>0\right)$ \cite{Dell2008}. As a result, the effective interatomic interaction $U_{1}^{eff1}$ (containing the dipolar interaction and
$s$-wave interaction) in Cr is less than the interatomic interaction $U_{2}$
in Rb, and both of them as well as the intercomponent interaction
are repulsive (see Table \ref{table:table1}). Then Cr pushes Rb out toward the edges of the trap, while the rubidium atoms also ``squeeze'' the chromium atoms, i.e., they act like a trap which enhances the trapping force on the chromium atoms and makes the peak density of Cr higher. In the one-dimensional case containing only $s$-wave scattering interactions, the criterion Eq. \eqref{eq:criterion} works perfectly \cite{Nav}. However, when
we increase $\varepsilon_{dd}$ (as in Fig. \ref{fig:Fig2} where
$\varepsilon_{dd}=0.1$), the anticipated phase separation doesn't
occur at the critical point. Instead, the two components admix
homogeneously [see Fig. \ref{fig:Fig2} (A)]. By calculating the
interactions' strengths, we know that the two interatomic
interactions and the intercomponent interaction have the same value (see Table \ref{table:table1}); thus the two-component Bose gas behaves like a one-component Bose gas since all particles have, so to speak, the same hard-core diameter. As a result, all the atoms are in a homogeneous mixture. Increasing $a_{12}$ slightly beyond the critical value of the intercomponent scattering length, we find that the two components are repelled by
each other. Since the intercomponent interaction is larger
than the two interatomic interactions as shown in Table
\ref{table:table1}, none of the atoms stays at the center of the trap
[see Fig. \ref{fig:Fig2} (B)].

\begin{figure}[b]\begin{center}
\includegraphics[width=1.0\linewidth]{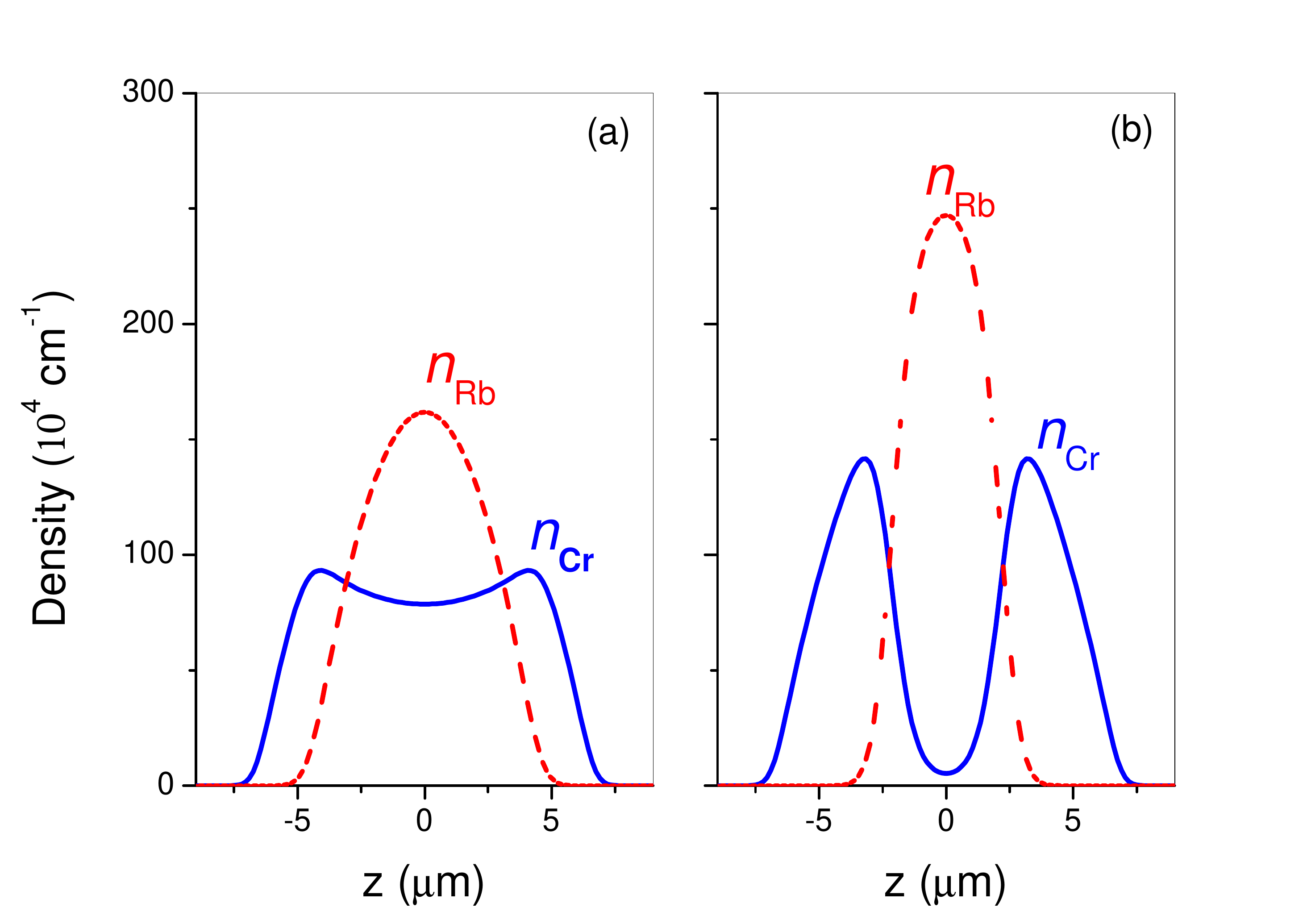}
\end{center}\vspace{-0.8cm}
\caption{\label{fig:Fig3} (Color online) Ground-state density profiles in the
two-component dipolar BEC for $\varepsilon_{dd}=0.5$ and different
values of the intercomponent scattering length: (A) $a_{12}=8$ nm; (B) $a_{12}=9.68$ nm (the critical point).}
\end{figure}

Increasing $\varepsilon_{dd}$ (as in the Fig.
\ref{fig:Fig3} where $\varepsilon_{dd}=0.5$), we find that the phase separation
is different from that shown in Fig. \ref{fig:Fig1}. At the critical point, the rubidium
atoms still occupy the center of the trap and their peak density becomes higher, while the chromium atoms
are pushed out instead [see Fig. \ref{fig:Fig3} (B)]. As we can see,
this is because the effective interatomic interaction of the
chromium BEC is now stronger than the interaction of the rubidium BEC
(see Table \ref{table:table1}), and the rubidium atoms get an extra trapping force from the chromium atoms. Based on this, we may control the
component which we intend to move out of the trap center to some extent.

\section{Quasi-two-dimensional Regime (infinite cigar)}\label{sec:twodregime}
We now consider the two-component dipolar cigar-shaped BEC in a
quasi-two-dimensional regime using a similar methodology to that for the
pancake shape. If we neglect the axial trapping
$\left(\omega_{iz}=0\right)$ and consider the BEC to be uniform
along $z$, then Eq. \eqref{eq:rwPhidd} reduces to the contactlike
form \cite{Dell2008}:
\begin{equation}
\Phi_{dd}=-U_1\varepsilon_{dd}n\left(\rho\right)=-U_1N_1\varepsilon_{dd}|\psi_1
\left(\rho\right)|^2.
\end{equation}
Then we can rewrite Eq. \eqref{eq:h1} and \eqref{eq:h2} as two
dimensionless coupled differential equations:

\begin{eqnarray}
H_1&=&-\frac{1}{2}\left(\frac{\partial^2}{\partial \rho^2}+\frac{1}{\rho}\frac{\partial}{\partial \rho}\right)+\frac{1}{2}\rho^2
+4\pi N_1a_1(1-\varepsilon_{dd})\times\nonumber\\&&|\psi_1(\rho)|^2+
\frac{1+a_m}{a_m}2\pi N_2a_{12}|\psi_2(\rho)|^2\\
H_2&=&-\frac{1}{2a_m}\left(\frac{\partial^2}{\partial \rho^2}+\frac{1}{\rho}\frac{\partial}{\partial \rho}\right)
+\frac{1}{2}a_{m}a_{\omega}^2\rho^2+\frac{4\pi N_2a_2}
{a_{m}}\times\nonumber\\&&|\psi_2 \left(\rho\right)|^2+
\frac{1+a_{m}}{a_{m}}2\pi N_1a_{12}|\psi_1\left(\rho\right)|^2,
\end{eqnarray}
with units of length and energy and $a_{m}=m_2/m_1$ as in Sec.
\ref{sec:onedregime}, while $a_{\omega}=\omega_{2\rho}/
\omega_{1\rho}$ is just different in form from that in Sec.
\ref{sec:onedregime}. Here, for simplicity and comparability, we assume that the values of
these two $a_{\omega}$ are equivalent, i.e.,
$\omega_{2z}/\omega_{1z}=\omega_{2\rho}/\omega_{1\rho}$.

Using a similar methodology, we apply boundary conditions
$\lim\limits_{\rho\rightarrow\infty}\psi_i\left(\rho\right)=0$ and
$\frac{\partial}{\partial\rho}\psi_i(\rho)|_{\rho=0}=0$, and the finite-difference
approximations \cite{finitediff}
\begin{eqnarray}
\frac{\partial^2\psi_i}{\partial\rho^2}=
\frac{\psi_i^{l+1}-2\psi_i^{l} +\psi_i^{l-1}}{h^2},~
\frac{\partial\psi_i}{\partial\rho}=
\frac{\psi_i^{l+1}-\psi_i^{l-1}}{2h}, \nonumber
\end{eqnarray}
then we can write $H_i$ as a nonsymmetric tridiagonal matrix containing diagonal elements $\varsigma_{i}^{l}$, subdiagonal elements $\varpi_{i}^{l}$, and superdiagonal elements $\vartheta_{i}^{l}$ in $H_{i}$, where
\begin{eqnarray}
&&\varsigma_{1}^{l}=2+2h^{2}\nu_{1}^{l},~
\varsigma_{2}^{l}=\frac{2}{a_{m}}+2h^{2}\nu_{2}^{l}, \nonumber\\
\nu_{1}^{l}&=&\frac{1}{2}\rho_{l}^{2}+4\pi N_{1}a_{1}\left(1-\varepsilon_{dd}\right)|\psi_{1}^{l}|^{2}+\frac{1+a_{m}}{a_{m}}2\pi N_{2}a_{12}|\psi_{2}^{l}|^{2}, \nonumber\\
\nu_{2}^{l}&=&\frac{1}{2}\rho_{l}^{2}+\frac{4\pi N_{2}a_{2}}{a_{m}}|\psi_{2}^{l}|^{2}+\frac{1+a_{m}}{a_{m}}2\pi N_{1}a_{12}|\psi_{1}^{l}|^{2}, \nonumber\\
&&\varpi_{1}^{l}=-\left(1-\frac{h}{2\rho_{l}}\right),~
\varpi_{2}^{l}=-\frac{1}{a_{m}}\left(1-\frac{h}{2\rho_{2}}\right), \nonumber\\
&&\vartheta_{1}^{1}=-2,~
\vartheta_{1}^{l\geq 2}=-\left(1+\frac{h}{2\rho_{l}}\right), \nonumber\\
&&\vartheta_{2}^{1}=-\frac{2}{a_{m}},~
\vartheta_{2}^{l\geq 2}=-\frac{1}{a_{m}}\left(1+\frac{h}{2\rho_{l}}\right), \nonumber\\
&&\rho_{l}=\left(l-1\right)h, \nonumber
\end{eqnarray}
with mesh length $h$ in the $\rho$ direction.
By diagonalizing these two nonsymmetric tridiagonal matrices, we can get
the ground state wave functions $\psi_i\left(\rho\right)$ and the density profiles of the dipolar BEC \cite{Pu}. Here, the critical value $a_{12}^{c}$ is obtained by
\begin{equation}
a_{12}^{c}=\sqrt{\frac{4a_1\left(1-\varepsilon_{dd}\right)a_2a_{m}}{\left(1+a_{m}
\right)^2}},
\end{equation}
arising from $U_1^{eff2}U_2-U_{12}^2=0$ with $U_1^{eff2}=U_1\left(1-\varepsilon_{dd}\right)$.

We assume that $\omega_{1\rho}=2\pi\times 160$ Hz, the particle
number is taken as $N_1=N_2=2000$, and $a_{m}a_{\omega}^2=1$ as well.

\begin{figure}[h]\begin{center}
\includegraphics[width=1.0\linewidth]{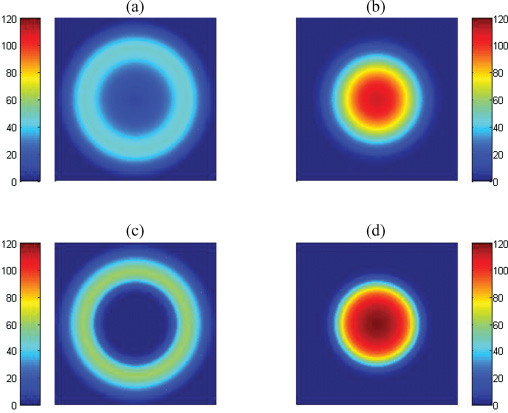}
\end{center}\vspace{-0.7cm}
\caption{\label{fig:Fig4} (Color online) Ground state density
profiles in the two-component dipolar BEC for
$\varepsilon_{dd}=-0.3$. The intercomponent scattering length
$a_{12}=5.5$ nm in (A) and (B), while $a_{12}=7.81$ nm (the critical point) in (C) and
(D). In this figure and Figs. \ref{fig:Fig5} and \ref{fig:Fig6}, (A) and (C) represent the chromium atoms, while (B) and (D)
represent the rubidium atoms. The numerical values of the units for density and length are respectively $1 \times 10^{8}$ cm$^{-2}$ and $1$ $\mu$m in our calculations. The field of view for (A)-(D) is $10 \times 10$ $\mu$m.}
\end{figure}

\begin{table}
\caption{\label{table:table2} Dimensionless values of interactions in quasi-two dimension}
\begin{tabular}{cccc}
\hline\hline
  $\varepsilon_{dd}$ & $U_{1}^{eff2}$ ($10^{-3}$) & $U_{12}$
  ($10^{-3}$) & $U_{2}$ ($10^{-3}$) \\ \hline
  -0.3 & 82 & \begin{tabular}{c}
                55 ($a_{12}=5.5$ nm) \\
                78 ($a_{12}^{c}=7.81$ nm) \\
              \end{tabular}
   & 75 \\
  0.1 & 57 & \begin{tabular}{c}
                30 ($a_{12}=3$ nm) \\
                65 ($a_{12}^{c}=6.5$ nm) \\
              \end{tabular}
   & 75 \\
  0.5 & 31 & \begin{tabular}{c}
                20 ($a_{12}=2$ nm) \\
                49 ($a_{12}^{c}=4.84$ nm) \\
              \end{tabular}
   & 75 \\
\hline\hline
\end{tabular}
\end{table}

\begin{figure}[b]\begin{center}
\includegraphics[width=1.0\linewidth]{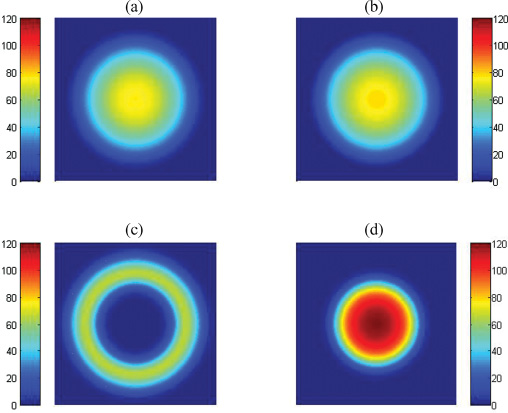}
\end{center}\vspace{-0.7cm}
\caption{\label{fig:Fig5} (Color online) Ground state density
profiles in the two-component dipolar BEC for $\varepsilon_{dd}
=0.1$. The intercomponent scattering length $a_{12}=3$ nm in (A) and
(B), while $a_{12}=6.5$ nm, the critical
point, in (C) and (D). The units are the same as in Fig. \ref{fig:Fig4}. The field of view for (A)-(D) is $10 \times 10$ $\mu$m.}
\end{figure}

In this case, the dipoles are predominantly end to end, and the
net dipolar interaction is repulsive (attractive) when
$\varepsilon_{dd}<0$ $\left(\varepsilon_{dd}>0\right)$ \cite{Dell2008}. This is
in contrast to the case in the quasi-one-dimensional regime. We then find that the
chromium atoms are squeezed out when $\varepsilon_{dd}=-0.3$ at the
critical point, and form a shell around the rubidium atoms which are situated at the center of the trap [see Fig. \ref{fig:Fig4} (C) and (D)]. The critical point of
$a_{12}$ in this case is not the same as in
the quasi-one-dimensional regime, but higher. We believe the cause is that the effective
interatomic interaction $U_{1}^{eff2}$ of Cr, which is repulsive,
becomes stronger than in the former case (see Tables
\ref{table:table1} and \ref{table:table2}). We cannot observe a similar
phase-separated state for $\varepsilon_{dd}=0.1$  as in Fig. \ref{fig:Fig2} (B), other than in the
quasi-one-dimensional regime. As shown in Fig. \ref{fig:Fig5}, the two components are mixed
homogeneously at a small value of $a_{12}$ (3 nm in Table \ref{table:table2}), and phase
separation occurs at the critical point.

\begin{figure}[t]\begin{center}
\includegraphics[width=1.0\linewidth]{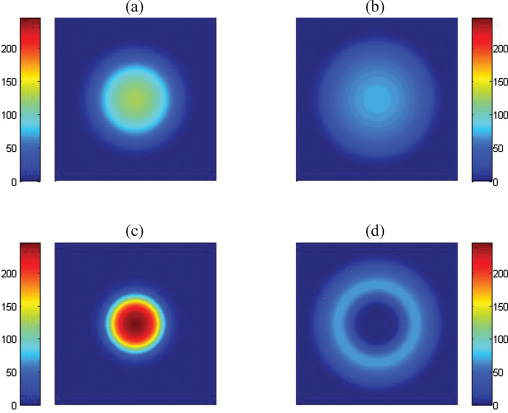}
\end{center}\vspace{-0.7cm}
\caption{\label{fig:Fig6} (Color online) Ground state density
profiles in the two-component dipolar BEC for $\varepsilon_{dd}=
0.5$. The intercomponent scattering length $a_{12}=2$ nm in (A) and
(B), while $a_{12}=4.84$ nm, the critical
point, in (C) and (D). The units are the same as in Fig. \ref{fig:Fig4}. The field of view for (A)-(D) is $10 \times 10$ $\mu$m.}
\end{figure}

\begin{figure}[b]\begin{center}
\includegraphics[width=1.0\linewidth]{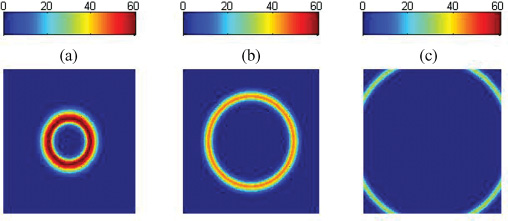}
\end{center}\vspace{-0.7cm}
\caption{\label{fig:Fig7} (Color online) Ground state density
profiles of component 1 (Cr). $N_{1}=2\times 10^{3}$,
$\varepsilon_{dd}=-0.3$, $a_{12}=a_{12}^{c}=7.81$. From (A) to (C),
$N_{2}=2\times 10^{3}$, $N_{2}=2\times 10^{4}$, $N_{2}=2\times
10^{5}$, respectively. The units are the same as in Fig. \ref{fig:Fig4}. The field of view for (A)-(C) is $20 \times 20$ $\mu$m.}
\end{figure}

\begin{figure}[t]\begin{center}
\includegraphics[width=1.0\linewidth]{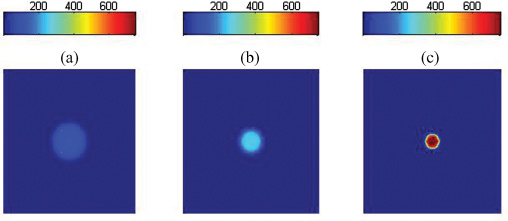}
\end{center}\vspace{-0.7cm}
\caption{\label{fig:Fig8} (Color online) Ground state density
profiles of component 2 (Rb). $N_{2}=2\times 10^{3}$,
$\varepsilon_{dd}=-0.3$, $a_{12}=a_{12}^{c}=7.81$. From (A) to (C),
$N_{1}=2\times 10^{3}$, $N_{1}=2\times 10^{4}$, $N_{1}=2\times
10^{5}$, respectively. The units are the same as in Fig. \ref{fig:Fig4}. The field of view for (A)-(C) is $20 \times 20$ $\mu$m.}
\end{figure}

On increasing $\varepsilon_{dd}$ to 0.5, we find that the attractive strength of the
dipolar interaction becomes very large, and the chromium atoms
begin to occupy the center of the trap, while the rubidium atoms
form a shell around chromium (see Fig. \ref{fig:Fig6}).
The phase-separated states for the same $\varepsilon_{dd}$ are totally
different between the quasi-one- and quasi-two-dimensional regimes. The
interactions are known to be the predominant factors which can affect
the ground-state density profile of a BEC, so different properties
of the dipolar interaction can make a difference to the phase
separations between the two regimes. For the same value of
$\varepsilon_{dd}$, the repulsion or attraction of the net dipolar
interaction is opposite in the two regimes, so the effective
interatomic interaction of Cr is totally different
between the two regimes (compare $U_{1}^{eff2}$ of Table \ref{table:table2} with $U_{1}^{eff1}$ of Table \ref{table:table1}). If $\varepsilon_{dd}$ is negative, the
repulsion of the effective interatomic interaction in the
quasi-two-dimensional regime is much stronger than in the
quasi-one-dimensional regime (see Tables \ref{table:table1} and \ref{table:table2}), so the rubidium atoms are held at the center of
the trap while the chromium atoms are forced to form a shell around
the center at the critical point of phase separation. The situation
for a relatively large positive $\varepsilon_{dd}$ is the opposite.

In addition to the strength of the interactions, the density profiles also
depend on the particle numbers $N_{1}$ and $N_{2}$ \cite{Pu}. Thus we do some calculations
where we hold constant the number of one component's atoms and change the other in the quasi-two-dimensional regime. Clearly,
the number of atoms plays a key role in the phase separation of a two-component BEC. In the case of $\varepsilon_{dd}=-0.3$, Cr forms a shell around Rb. When the number of Cr atoms is fixed, they are squeezed further out as the number of Rb atoms increases (See Fig. \ref{fig:Fig7}). If we fix the number of Rb atoms instead, one can see that the rubidium atoms are compressed further as the number of Cr increases (See Fig. \ref{fig:Fig8}). Finally, we would like to mention that the situation of the quasi-one-dimensional regime is similar to this case herein.

\section{Conclusion}\label{sec:conclusion}
In the present work, we have analyzed the phase separation of a
two-component dipolar BEC in the quasi-one- and quasi-two-dimensional
regimes, as a natural parameter of the system (the intercomponent
interaction strength) was varied. Our analysis was presented for the
case of a two-component Bose-Einstein condensate trapped in a
harmonic-oscillator potential.

We reduced the long-range dipole-dipole interaction to a contactlike
form, when a dipole moment is aligned along the $z$ direction, and
then used a numerical method to solve two coupled nonlinear
Gross-Pitaevskii equations in the low-dimensional regime. We were able to
elucidate the phase separation condition for the intercomponent
coupling parameter as the dimensionless dipolar interaction strength
parameter $\varepsilon_{dd}$ was varied.

Both repulsive and attractive dipolar interactions have been taken into consideration.
Among all the phase-separated states, $\varepsilon_{dd}$ plays a
significant role. By modulation of its strength, the two components might alternately stay in or
be forced to be outside the center of the trap. In addition to the interactions, we considered the
dependence of the density profiles on the number of atoms as well.

We also described the differences between the quasi-one- and
quasi-two-dimensional regimes in detail. Due to the anisotropic property of
the dipole-dipole interaction, phase separation in the axial
direction differs from that in the radial direction.

A natural extension of this work is to try to generalize the ideas
presented herein to higher dimensions or larger number of
components (i.e., spinors). Especially in the former case,
more complicated pattern formations on the interface, such as
Rosensweig hexagonal peaks and a labyrinthine pattern, have been
observed \cite{Ueda2009}. Such studies are presently in progress and
will be presented in future work.

\section{Acknowledgement}\label{sec:ack}
J. B. L. is grateful to Yue Yu for his encouragement. We
would also like to thank Su Yi for comments. This work is supported
by the Natural Science Foundation of Jiangsu Province (NSFJS, Grant No. BK2010499) and MOE of China (Grant No. 20070287062).


\begin{thebibliography}{1000}

\bibitem{Pfau2002} S. Giovanazzi, A. G\"{o}rlitz, and T. Pfau, Phys. Rev. Lett. {\bf 89}, 130401 (2002).
\bibitem{Pfau2005} A. Griesmaier, J. Werner, S. Hensler, J. Stuhler, T. Pfau, Phys. Rev. Lett. {\bf 94}, 160401 (2005); J. Stuhler, A.
Griesmaier, T. Koch, M. Fattori, T. Pfau, S. Giovanazzi, P. Pedri,
L. Santos, {\it ibid.} {\bf 95}, 150406 (2005); M. Fattori, T. Koch,
S. Goetz, A. Griesmaier, S. Hensler, J. Stuhler, T. Pfau,
 Nature Phys. {\bf 2}, 765 (2006).
\bibitem{Pfau2006} A. Griesmaier, J. Stuhler, T. Koch, M. Fattori, T. Pfau,
  S. Giovanazzi, Phys. Rev. Lett. {\bf 97}, 250402 (2006).
\bibitem{Pfau2007} T. Lahaye, T. Koch, B. Fr\"{o}lich, M. Fattori, J. Metz, A. Griesmaier,
S. Giovanazzi, T. Pfau, Nature {\bf 448}, 672 (2007). T. Koch,
T. Lahaye, J. Metz, B. Fr\"{o}lich, A. Griesmaier, T. Pfau,
 Nature Phys. {\bf 4}, 218 (2008); T. Lahaye, J. Metz, B. Fr\"{o}lich, T. Koch, M. Meister,
 A. Griesmaier, T. Pfau, H. Saito, Y. Kawaguchi, M. Ueda, Phys. Rev. Lett. {\bf 101},
 080401 (2008).
\bibitem{Pfau2000} K. G\'{o}ral, K. Rz\c{a}\.{z}ewski, and T. Pfau, Phys. Rev. A {\bf 61},
 051601 (2000). L. Santos, G.V. Shlyapnikov, P. Zoller, and M. Lewenstein, Phys. Rev.
Lett. {\bf 85}, 1791 (2000).
\bibitem{Dell2002} S. Giovanazzi, D. O'Dell, and G. Kurizki, Phys. Rev. Lett. {\bf 88},
130402 (2002). K. G\'{o}ral, L. Santos, and M. Lewenstein, {\it
ibid} {\bf 88}, 170406 (2002).
\bibitem{Yi2000} S. Yi and L. You, Phys. Rev. A {\bf 61}, 041604(R) (2000).
\bibitem{Yi200102} S. Yi and L. You, Phys. Rev. A {\bf 63}, 053607 (2001); S. Yi and L.
You, {\it ibid.} {\bf 66}, 013607 (2002).
\bibitem{Santos2002_2} K. G\'{o}ral and L. Santos, Phys. Rev. A {\bf 66}, 023613 (2002).
\bibitem{Lahaye2009} T. Lahaye, C. Menotti, L. Santos, M. Lewenstein, and T. Pfau,
Rep. Prog. Phys. {\bf 72}, 126401 (2009).
\bibitem{Ho} Tin-Lun Ho and V. B. Shenoy, Phys. Rev. Lett. {\bf 77}, 3276 (1996); E.
Timmermans, {\it ibid.} {\bf 81}, 5718 (1998).
\bibitem{Pu} H. Pu and N. P. Bigelow, Phys. Rev. Lett. {\bf 80}, 1130 (1998).
\bibitem{Cornell1997} C.J. Myatt, E.A. Burt, R.W. Ghrist, E.A. Cornell, and C.E. Wieman,
Phys. Rev. Lett. {\bf 78}, 586 (1997); D.M. Stamper-Kurn, M.R.
Andrews, A.P. Chikkatur, S. Inouye, H.-J. Miesner, J. Stenger, and
W. Ketterle, Phys. Rev. Lett. {\bf 80}, 2027
  (1998); D.S. Hall, M.R. Matthews, J.R. Ensher, C.E. Wieman, and E.A. Cornell,
   Phys. Rev. Lett. {\bf 81}, 1539 (1998).
\bibitem{Cornell2004} V. Schweikhard, I. Coddington, P. Engels, S. Tung, and E.A. Cornell,
 Phys. Rev. Lett. {\bf 93}, 210403 (2004); K.M. Mertes, J.W. Merrill,
 R. Carretero-Gonz\'{a}lez, D.J. Frantzeskakis, P.G. Kevrekidis, and D.S. Hall, {\it ibid.}
 {\bf 99}, 190402 (2007); S.B. Papp, J.M. Pino, and C.E. Wieman, {\it ibid.} {\bf 101}, 040402
 (2008).
\bibitem{Ueda2009} Hiroki Saito, Yuki Kawaguchi, and Masahito Ueda, Phys. Rev. Lett.
{\bf 102}, 230403 (2009).
\bibitem{Malomed2010} Goran Gligori\'{c}, Aleksandra Maluckov, Milutin Stepi\'{c},
Ljup\v{c}o Had\v{z}ievski, and Boris A. Malomed, Phys. Rev. A
{\bf  82}, 033624 (2010).
\bibitem{Pethick} C. J. Pethick and H. Smith, {\it Bose-Einstein Condensation in Dilute Gases}, 2nd ed. (Cambridge University Press, Cambridge, U.K., 2008).
\bibitem{Dell20040507} D. O'Dell, S. Giovanazzi, and C. Eberlein, Phys. Rev. Lett.
{\bf 92}, 250401 (2004); C. Eberlein, S. Giovanazzi, and D. O'Dell, Phys. Rev. A
{\bf 71}, 033618 (2005); D. O'Dell and C. Eberlein, {\it ibid.} {\bf 75}, 013604 (2007).
\bibitem{Dell2008} N.G. Parker and D. O'Dell, Phys. Rev. A {\bf 78}, 041601(R) (2008).
\bibitem{finitediff} J.W. Thomas, Numerical Partial Differential Equations: Finite Difference Methods, Springer, 1995.
\bibitem{Chui1998} P. Ao and S. T. Chui, Phys. Rev. A {\bf 58}, 4836 (1998).
\bibitem{Feshbach} M. Theis, G. Thalhammer, K. Winkler, M. Hellwig, G. Ruff, R. Grimm, and J. H. Denschlag, Phys. Rev. Lett. {\bf 93}, 123001 (2004).
\bibitem{Nav} R. Navarro {\it et al}., Phys. Rev. A {\bf 80}, 023613 (2009); S. Gautam and D. Angom, J. Phys. B: At. Mol. Opt. Phys. {\bf 44} 025302 (2011).
\end{thebibliography}
\end{document}